\documentclass[twocolumn,pra,showpacs,superscriptaddress]{revtex4-1}
\usepackage{amssymb}
\usepackage{amsmath}
\usepackage{graphicx}
\usepackage{subfigure}
\usepackage{natbib}
\usepackage{epsfig}
\usepackage{amsfonts}
\usepackage{mathrsfs}
\usepackage{CJK}
\usepackage{xcolor}
\usepackage[toc,page,title,titletoc,header]{appendix}

\begin{document}

\title{Multi-mode effects in cavity QED based on a one-dimensional
  cavity array}

\author{Wei Zhu}

\affiliation{Institute of Physics, Beijing National Laboratory for
  Condensed Matter Physics, Chinese Academy of Sciences,Beijing
  100190, China}

\author{Z. H. Wang}

\affiliation{Beijing Computational Science Research Center, Beijing
  100084, China}

\author{D. L. Zhou}

\email{zhoudl72@iphy.ac.cn}

\affiliation{Institute of Physics, Beijing National Laboratory for
  Condensed Matter Physics, Chinese Academy of Sciences,Beijing
  100190, China}

\date{\today}

\begin{abstract}
  We present a microscopic model of cavity quantum electrodynamics
  (QED) based on a one-dimensional coupled cavity array (1D CCA),
  where a super cavity (SC) is composed by a segment of the 1D CCA
  with relatively smaller couplings with the outsides. The
  single-photon scattering problem for the SC empty or with a
  two-level atom in is investigated. We obtain the exact theoretical
  result on the transmission rate for our system, which predicts the
  transmission peaks shall appear near the eigen-energies of the SC.
  Our numerical results further prove that the SC is a well-defined
  multi-mode cavity. When a two-level atom resonant with the SC
  locates at the antinode of the resonant mode, the transmission
  spectrum shows a clear sign of vacuum Rabi splitting as expected.
  However, when the atom locates at the node of the resonant mode, we
  observe a deep valley in the transmission peak, which can be
  explained by the destructive interference of two transmission
  channels, one is the resonant mode, while the other is arising from
  the atom coupling with the non-resonance modes. The effect of
  non-resonance modes on vacuum Rabi splitting is also analyzed.
\end{abstract}

\maketitle

\section{Introduction}

Cavity QED, the study of the interaction between atoms and the
quantized electromagnetic fields in a micro-cavity, has been one of
the central research areas in both quantum optics and quantum
information since the pioneering work of Purcell~\cite{1a}. In a
single cavity with one atom or multi-atoms, the hallmark phenomena,
such as vacuum Rabi splitting~\cite{1b}, Rabi
oscillation~\cite{brune}, collective Lamb shift~\cite{RR}, and
electromagnetically induced transparency~\cite{MM}, have been
successfully observed.

With the rapid development of experimental technologies in recent
years, the system of CCA with atoms embedded in has aroused
significant attentions. It is a promising test bed which is widely
used in various areas and also a building block for important quantum
devices. In quantum simulation, many important phenomena in condensed
matter have been successfully observed on this platform, such as
Mott-superfluid transitions~\cite{4,5,6} and some topological
effects~\cite{7,8,9,10}.

CCA also shows its application in controlling single-photon which is
of essential importance in quantum optics and quantum information. One
of the pioneering work~\cite{1} focuses on a 1D CCA doped with a
two-level system, and shows that the controllable system can behave as
a quantum switch for the coherent transport of single photon.
Furthermore, the single-photon scattering for 1D CCA with a pair of
two-level atoms or with three-level atoms has also been
discussed~\cite{3}. Aside from the 1D CCA, single-photon scattering
for 2D CCA is also important for its promising application in quantum
networks~\cite{3a}. In addition, CCA may be introduced in more
research areas, e.g., a recent paper designed an experiment on CCA to
explore the basic principle in quantum mechanics~\cite{3b}.

In this paper, we use CCA to investigate the basic problem in cavity
QED. Based on 1D CCA, we propose a new real space cavity model with a
SC as our cavity. The characteristics of SC empty or with a two-level
atom doped in is explored by studying the single-photon scattering
problem. Applying the discrete coordination scattering
equation~\cite{1} to the case of empty SC, we give the transmission
spectrum numerically and analytically prove that the transmission
peaks shall appear near the eigenvalues of the empty SC system. These
transmission peaks with non-zero width imply that the empty SC can be
considered as a well defined multi-mode cavity with dissipation.

In particular, we consider the variation of the single-photon
transmission due to a two-level atom which is near resonant with one
of the SC modes. When the atom is located at the antinode of the
resonant mode, we observe the vacuum Rabi splitting in the
transmission spectrum. However, when the two-level atom is located at
the node of the resonant mode, i.e., it does not interact with the
resonant mode, an obvious valley in the transmission spectrum is
observed, which is further explained by the destructive interference
of two transmission channels, one is provided by the resonant mode of
the SC, while the other results from the two-level atom which couples
to the non-resonance modes of the SC.

The rest of the paper is organized as follows. In Sec.~\ref{sec:2}, we
introduce the theoretical model of our system and give the analytical
result of the single-photon transmission rate. In Sec.~\ref{sec:3}, we
study the single-photon scattering on the empty SC, which shows the
characteristic features of the SC. In Sec.~\ref{sec:4}, we detailed
investigate how a two-level atom essentially change the transmission
spectrum, especially focus on the effect of non-resonance modes of the
SC. In Sec.~\ref{sec:4c}, we introduce the two level approximation of
the SC system to physically explain the transmission valley as the
destructive interference between the two transmission channels
assisted by the two levels respectively. In Sec.~\ref{sec:5}, we
briefly discuss the experimental feasibility of the theoretical
predictions and give a summary of our results.

\section{Model and theoretical results \label{sec:2}}

\begin{figure}[!htbp]
  \includegraphics[width=8cm]{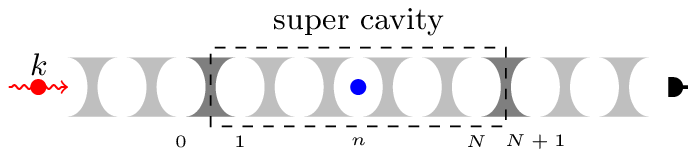}
  \caption{(Color online). Schematic configuration of the
    single-photon scattering problem for the 1D CCA model. A single
    photon (filled red circle) with the wave vector $k$ injects from
    the left side of the super cavity composed of $N$ cavities, which
    is formed by a relatively small coupling strength $\eta$ with the
    outside cavities. A two-level atom (filled blue circle) is in the
    $n$-th cavity of the SC. The transmission spectrum is measured by
    the detector on the right side of the SC. Here we take $N=5$ and
    $n=3$.}
  \label{fig:1}
\end{figure}

The system we consider is composed by a two-level atom interacting
with the $n$-th cavity in a 1D coupled single-mode cavity array with
infinite length, which is shown in Fig.~\ref{fig:1}. The hopping
strength between neighboring cavities $l$ and $l+1$ is $\xi$ for
$l\notin\{0,N\}$, and the hopping strength is $\eta$ for
$l\in\{0,N\}$, which is much less than $\xi$. In such setting,
the cavities between $1$ and $N$ forms a secondary
cavity, which we will call super cavity thereafter. Meanwhile,
the two-level atom is required to be inside the super cavity, i.e.
$1\le{n}\le{N}$.

A tight-bonding model including five parts is introduced to describe
the system
\begin{equation}
  H = H_{S} + H_{L} + H_{R} + H_{SL} + H_{SR},
\end{equation}
where
\begin{eqnarray}
  \label{eq:1}
  H_{S} & = & \sum^{N}_{j=1} \omega_{c}a^{\dag}_{j}a_{j} 
  + \sum^{N}_{j=2} \xi(a^{\dag}_{j-1}a_{j}+a^{\dag}_{j}a_{j-1}) \nonumber\\
  & & + \omega_{a}|e\rangle\langle e|
  + g (a^{\dag}_{n}\sigma^{-}+\mathrm {H.c.}), \label{aj}\\
  H_{L} & = & \sum^{0}_{j=-\infty} [ \omega_{c}a^{\dag}_{j}a_{j}
  + \xi (a^{\dag}_{j-1}a_{j}+a^{\dag}_{j}a_{j-1}) ], \\
  H_{R} & = & \sum^{\infty}_{j=N+1} [ \omega_{c}a^{\dag}_{j}a_{j} +
  \xi(a^{\dag}_{j}a_{j+1}+a^{\dag}_{j}a_{j+1}) ],\\
  H_{SL} & = & \eta(a^{\dag}_{0}a_{1} + \mathrm{H.c.}),\\
  H_{SR} & = & \eta(a^{\dag}_{N}a_{N+1}+ \mathrm{H.c.}).
\end{eqnarray}
Here $H_{S}$ describes the SC system, $H_{L}$ ($H_{R}$) describes the
left (right) channel that is formed by the segment of the cavity array
left (right) to the SC, and $H_{SL}$ ($H_{SR}$) describes the
interaction between the SC with the outsides. $a^{\dag}_{j}$ ($a_{j}$) is the
photon creation (annihilation) operator for the $j$-th single-mode
cavity, $|e\rangle$ ($|g\rangle$) is the excited (ground) state of the
atom, and $\sigma^{-}$($\sigma^{+}$) is the atomic lowering (raising)
operator. $\omega_{c}$ is the intrinsic frequency of each single-mode
cavity, $\omega_{a}$ is the transition frequency of the two-level
atom, and $g$ is the coupling strength between the cavity and the
atom. $\xi$ is the coupling strength between neighboring cavities in
the SC, the left channel and the right channel, while $\eta$ is the
coupling strength of neighboring cavities between them. In addition,
we require that $\eta\ll\xi$, and set $\hbar=1$ throughout this paper.

The basic task here is to investigate the single photon scattering
problem. When a single photon with wave vector $k$ injects from the
left channel towards the SC system, what is the transmission spectrum
obtained in the right channel?

Now the scattering state can be expanded as
\begin{equation}
  |\Psi_{k}\rangle = |\phi_{k}\rangle + r |\phi_{k}^{\ast}\rangle
  + t |\vartheta_{k}\rangle + \sum^{N}_{j=1} d_{j} |j\rangle + \lambda |e\rangle,
  \label{state}
\end{equation}
where
\begin{eqnarray}
  |\phi_{k}\rangle & = & \sum^{0}_{j=-\infty}e^{ikj}|j\rangle, \\
  |\vartheta_{k}\rangle & = & \sum^{+\infty}_{j=N+1}e^{ikj}|j\rangle,
\end{eqnarray}
with $|j\rangle=a_{j}^{\dagger}|G\rangle$ and
$|e\rangle=\sigma^{+}|G\rangle$. Here, $|G\rangle=|vac;g\rangle$
represents the state with all the cavities in their vacuum
states while the atom in the ground state. In Eq.~(\ref{state}),
the coefficients $r$ and $t$ are the reflection and transmission
amplitudes respectively, $d_{j}$ is the probability amplitude for
finding the photon in the $j$-th cavity, and $\lambda$ is the
probability amplitude for the atom in the excited state. The
scattering state satisfies the stationary Schr\"{o}dinger equation
\begin{equation}
  \label{eq:2}
  H|\Psi_{k}\rangle=E_{k}|\Psi_{k}\rangle.
\end{equation}

In general, the transmission amplitude $t$ is completely determined by
Eq.\eqref{eq:2}, and the transmission rate $T={|t|}^{2}$ can be
obtained. One of our main theoretical results is
\begin{equation}
  \label{eq:3}
  t =\frac { i e^{-i(k+\pi)(N+1)} 2 \gamma^{2} \sin k } { e^{-2ik} |A_{N}^{n}| +
    \gamma^{2} e^{-ik} (|A_{N-1}^{n}|+|A_{N-1}^{n-1}|) + \gamma^{4} |A_{N-2}^{n}|},
\end{equation}
where $\gamma=\eta/\xi$, and $A_{N}^{n} = [H_{S}(N,n)-E_{k}]/\xi$,
with the dispersion relation $E_{k}=\omega_{c}+2\xi\cos k$ (the wave
vector $k$ is dimensionless by setting the distance between arbitrary
two neighbouring cavities as unit), $H_{S}(N,n)$ is the Hamiltonian of
the SC system with $N$ cavities and the two-level atom in the $n$-th
cavity, $|A_{N}^{n}|$ is the determinant of $A_{N}^{n}$. The detailed
derivation of Eq.\eqref{eq:3} is given in the Appendix.

Since $\gamma$ is a small parameter, the transmission peaks occur only
when $|A_{N}^{n}|$ is small at least in the order of $\gamma^{2}$. In
addition, $|A_{N}^{n}|=0$ only when $E_{k}$ is the eigen-energy of the
SC system. Therefore, the necessary condition to observe the
transmission peaks is that $E_{k}$ is near resonant with the
eigen-modes of the SC system.

\section{Single photon scattering with empty super
  cavity \label{sec:3}}

As the first step, we study the transmission spectrum for the SC
without the two-level atom, i.e., an empty SC. As is well known, the
eigenvalues and eigenstates of the empty SC are 
\begin{eqnarray}
  \nu_{m} & = & \omega_{c} + 2\xi\cos(j\theta_{m}), \\
  |\Phi_{m}\rangle & = & \sqrt{\frac{2}{N+1}} \sum^{N}_{j=1}\sin(j \theta_{m})
  |j\rangle,
\end{eqnarray}
where $\theta_{m}=\frac{m \pi}{N+1}$ with $m$ being any integer between $1$ and $N$.

\begin{figure}[!htbp]
  \includegraphics[width=8cm]{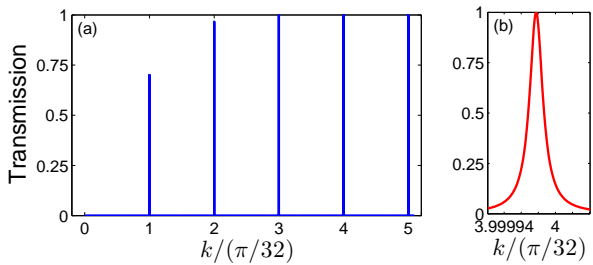}
  \caption{(Color online). (a) The transmission rate vs the incident
    wave vector $k$ for the empty SC. Only the first five transmission
    peaks are shown. (b) Zoom in of the $4$-th transmission peak. Here
    we choose the parameters $N=31$, $\xi=1$, and $\eta=0.01$.}
  \label{empty}
\end{figure}

Then the peaks in the transmission spectrum shall appear near the
resonance $E_{k}=\nu_{m}$, i.e. $k=\theta_{m}$, which is numerically
demonstrated in Fig.~\ref{empty}(a). In the figure, we only show the
first five peaks, where the fourth peak is zoomed in in
Fig.~\ref{empty}(b). It is apparent that each peak has a non-zero
width which means the corresponding electromagnetic mode in the SC
system has finite lifetime due to the dissipation arising from the
coupling with the left and right channels. In other words, the left
and right channels act not only as the carriers of the scattering
waves, but also as the dissipation reservoirs of the SC.

As demonstrated above, the numerical results show that the SC is
essentially an $N$-mode cavity, and every eigen-mode of the SC has a
relatively long life time. In what follows, we will study how a single
two-level atom that couples with the SC can dramatically change the
transmission spectrum, and investigate the effects induced by the
multi-mode cavity fields.

\section{Single photon scattering with one atom in the super
  cavity \label{sec:4}}

When a two-level atom interacts with the $n$-th cavity, it is
convenient to rewrite the Hamiltonian $H_{S}$ in the eigen-modes of
the SC as
\begin{equation}
  \label{eq:5}
  H_{S} = \sum_{k} \nu_{k} b_{k}^{\dagger} b_{k} + \omega_{a} |e\rangle \langle e|
  + \sum_{k} g_{k} (b_{k}^{\dagger} \sigma_{-} + b_{k} \sigma_{+}),
\end{equation}
where
\begin{eqnarray}
  b^{\dag}_{k} & = & \sqrt{\frac{2}{N+1}} \sum^{N}_{j=1} \sin(j\theta_{k})
  a^{\dag}_{j},\\
  g_{k} & = & g \sqrt{\frac{2}{N+1}} \sin(j\theta_{k}).
\end{eqnarray}
Note that the coupling strength $g_{k}$ depends on the location of
the atom and the length of the SC. It is the standard model of cavity
QED in the multi-mode setting.

When the frequency of the atom is near resonant with a pre-selection
$k^{\ast}$th mode of the SC, we may adopt the single mode
approximation for the SC. Then the two relative eigen-energy levels
are
\begin{equation}
  E_{k\pm} = \frac{\nu_{k^{\ast}}+\omega_{a}}{2} \pm \frac{\Delta_{k^{\ast}}}{2},
\end{equation}
where the vacuum Rabi splitting is
\begin{equation}
  \Delta_{k^{\ast}} = \sqrt{(\nu_{k^{\ast}} - \omega_{a})^{2} + 4 g_{k^{\ast}}^{2}}.
\end{equation}

\begin{figure}[!htbp]
  \includegraphics[width=8cm]{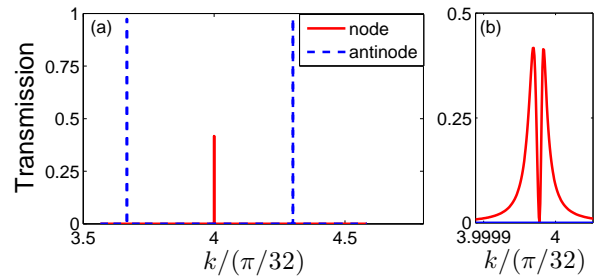}
  \caption{(Color online). (a) Transmission rates vs the incident
    wave vector (just around the resonant mode). Blue dashed and red
    solid line each stands for the atom at the antinode (n=12) or
    node (n=8) of the mode. (b) Zoom in of of the red peak. Here the
    atom is near resonant with the $4$-th mode of the SC with 
    $\omega_{a}=1.847760755\xi$, and $g=0.1$.}
  \label{resonant}
\end{figure}

Now we numerically check the validness of the above single mode
approximation. First, we plot the transmission spectrum when the
two-level atom is located at the node and antinode of the resonant
mode of the SC, which is shown in Fig.~\ref{resonant} (a). When the
atom is located at the antinode of the resonant mode, the obvious
vacuum Rabi splitting appears. When the atom is located at the node,
i.e., the two-level atom does not interact with the resonant mode of
the SC, no vacuum Rabi splitting is observed as expected. However, the
peak is lower than $1/2$ in the latter case. If the single mode
approximation is valid, then the two-level atom without interacting
with the resonant mode of the SC will not affect the transmission
rate. In other words, it is expected to be similar to the case shown
in Fig.~\ref{empty}(b). Thus we plot the zoom in for the case of the
atom at the node as shown in Fig.~\ref{resonant}(b). Obviously, the
transmission spectrum shown in Fig.~\ref{empty} (b) exhibits an
obvious valley exactly at the resonant mode of the SC, which is
essentially different from Fig.~\ref{one}(b).

\begin{figure}[!htbp]
  \includegraphics[width=8cm]{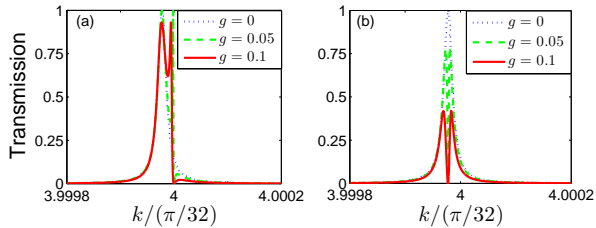}
  \caption{(Color online). Transmission spectrum for the atom at the
    node of the resonant mode with $g=0$ (blue dotted line), $g=0.05$
    (green dashed line) and $g=0.1$ (red solid line). (a) $\omega_{a}=\nu_{4}$ . 
    (b) $\omega_{a}=1.847760755\xi$.
  }
  \label{tunnel}
\end{figure}

To investigate the physics underlying the transmission spectrum shown
in Fig.~\ref{resonant}(b), the frequency of the atom is tuned to be
resonant with the mode while keeping the atom at the node, the
transmission spectrum is given for different coupling strength $g$ as
shown in Fig.~\ref{tunnel}(a). When $g=0$, we recover the transmission
spectrum shown in Fig.~\ref{empty}(b). For $g$ is not $0$, a second
peak appears near the frequency of the atom. Since the atom does not
couple with the resonant mode, the peak must be originated from the
atom coupling with the non-resonance modes. We assert that due to the
different influence of the outsides, the transmission spectrum for
each channel independently is partially overlap at resonant condition
which leads to the above mentioned transmission spectrum as shown in
next section. By increasing $g$, the transmission peak becomes wider
as expected. The numerical result in Fig.~\ref{tunnel}(a) clearly
shows that in the case of the atom located at the node of the resonant
mode, two channels exist for the photon transmitting through the SC,
one is the resonant cavity while the other is result from the atom
coupling with the non-resonance modes.

Further in Fig.~\ref{tunnel}(b), we tune the frequency of the atom
so that the transmission peaks for the two channels coincides as shown 
in next section. However, we find that the single
photon is completely reflected at the original transmission peak of the
resonant mode, which implies the transmission amplitudes through the two channels
interfere destructively. The appearance of the transmission valley
comes from the different widths of the transmission peaks from the
two channels. Moreover, the widths of the valleys are determined by
the coupling strength $g$, and it further confirms the existence of
quantum interference. A more physical explanation of the transmission
spectrum is given in the next section to intuitively show the mechanism
behind the transmission dip.

\begin{figure}[!htbp]
  \includegraphics[width=6cm]{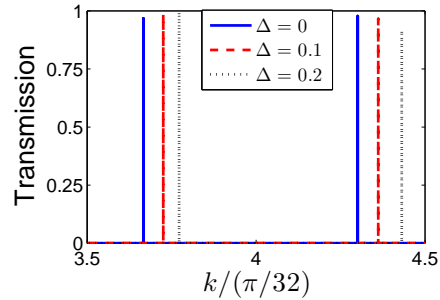}
  \caption{(Color online). Transmission spectrum for different
    detunings when the atom located in the antinode. Here we give the
    results for $\Delta=\omega_{a}-\nu_{4}=0$ (blue solid line),
    $\Delta=0.01\xi$ (red dashed line), and $\Delta=0.02\xi$ (black dotted
    line).}
  \label{detune}
\end{figure}

As discussed above, the non-resonance modes dramatically
change the transmission spectrum when the atom is located at the
node, so a natural problem is how the non-resonance modes affect the
transmission spectrum when the two-level atom is located at the
antinode of the resonant mode. To this end, we plot the transmission
spectrum for different detuning when the atom is located at the
antinode as shown in Fig.~\ref{detune}. We show that, by tuning the
frequency of the atom higher than the resonant mode, the peak of high
frequency will move away faster than the one with low frequency, and
the opposite behavior can be seen if tuning the atom frequency lower.
Obviously, this behavior can not be explained by the single mode
approximation, and must be resorted to the effects of non-resonance
modes of the SC. Notice that the similar phenomenon is mentioned in
Ref.~\cite{11}.

\section{Physical explanation of the transmission valley in the
  two level approximation}
\label{sec:4c}

Now let us give a more detailed and physical explanation about the
transmission valley shown in Fig~\ref{tunnel}. As we know, the energy levels of
the SC that are near resonant with the energy of the incoming photon
dominates the photon transmission through the SC. In the cases that
the transmission valley occurs, there are two energy levels of the SC
that are near resonant with the energy of the incoming photon, one is
the near resonant mode of the SC that does not interact with the atom,
the other is the atomic excited state dressed by the non-resonance
modes. So it is reasonable to maintain only these two levels in the
Hamiltonian $H_{S}$, which is called the two level approximation. In
the two level approximation, the Hamiltonian of our system can be
written as $H=H_S+H_L+H_R+H_{SL}+H_{SR}$ with $H_L$ and $H_R$ the same
as in the exact model and
\begin{eqnarray}
  \label{eq:1}
  H_{S} & = & \nu_{m} |\psi_{m}\rangle \langle \psi_{m}|+\omega_{A}
  |\varphi_{m}\rangle \langle \varphi_{m}|, \\
  H_{SL} & = & \eta\left[(\alpha_{1} |0\rangle \langle
    \psi_{m}|+\beta_{1} |0\rangle \langle  \varphi_{m}|) + \mathrm {H.c.}\right],\\
  H_{SR} & = & \eta\left[(\alpha_{2} |N+1\rangle\langle
    \psi_{m}|+\beta_{2} |N+1\rangle  \langle \varphi_{m}|)+ \mathrm {H.c.}\right],
\end{eqnarray}
where $|\psi_{m}\rangle=b^{\dag}_{m}|G\rangle$
$\alpha_{1}=\langle 1|\psi_{m}\rangle$, $\beta_{1}=\langle 1|\varphi_{m}\rangle$,
$\alpha_{2}=\langle N|\psi_{m}\rangle$, and $\beta_{2}=\langle N|\varphi_{m}\rangle$.
$\omega_{A}$ is the eigen-energy of the atomic state.

Now the scattering state can be expanded as
\begin{equation}
  |\Psi_{k}\rangle =|\varphi_{k}\rangle + r |\varphi_{k}^{\ast}\rangle + t
  |\vartheta_{k}\rangle
  + \mu |\psi_{m}\rangle + \zeta |\varphi_{m}\rangle
\end{equation}
with $\mu$ and $\zeta$ being the excitation amplitudes of the modes
$|\psi_{m}\rangle$ and $|\varphi_{m}\rangle$, respectively.

The stationary Schr\"{o}dinger equation
$H|\Psi_{k}\rangle=E_{k}|\Psi_{k}\rangle$ results in the following set
of scattering equations
\begin{eqnarray}
  \omega_{c}r^{'}+\xi(e^{-ik}+re^{ik})+\eta(\alpha_{1} \mu + \beta_{1}
  \zeta ) & = & E_{k}r^{'} \\
  \eta[r^{'}\alpha_{1}+t^{'}\alpha_{2}] + \nu_{m}\mu& = &
  E_{k}\mu \label{a1} \\
  \eta[r^{'}\beta_{1}+t^{'}\beta_{2}] + \omega_{A}\zeta& = &
  E_{k}\zeta \\
  \omega_{c}t^{'}+\eta(\alpha_{2}\mu + \beta_{2}\zeta)+\xi t^{'}e^{ik}
  & = & E_{k}t^{'}\label{a2}
\end{eqnarray}
and the transmission rate $T=|t|^2$ can be determined. Here, we set
$t^{'}=t e^{ik(N+1)}$ and $r^{'}=1+r$.

From Eq.~\eqref{a2}, the condition for the perfect reflection is
\begin{equation}
  \label{eq:int}
  \alpha_{2}\mu + \beta_{2}\nu=0.
\end{equation}
Eq.~\eqref{eq:int} can be understood as follows. When the three levels
$|\psi_{m}\rangle$, $|\varphi_{m}\rangle$, and $|N+1\rangle$ are considered, the
state $\mu |\psi_{m}\rangle + \zeta |\varphi_{m}\rangle$ is the dark
state relative to $|N+1\rangle$ as in the EIT setting. This shows the
interference mechanism behind the transmission valley and is the
central point to understand the phenomenon.

\subsection{Analytical and numerical results for $\omega_{a}=\nu_{m}$}
\label{sec:analyt-results-omeg}

When the atom is resonant with the $m$-th mode of the cavity
($\omega_{a}=\nu_{m}$), the state $|\varphi_{m}\rangle$ with eigen-energy
$\omega_{A}=\omega_{a}$ can be analytically expressed as
\begin{equation}
  |\varphi_{m}\rangle = c_{m}^{\dag}|G\rangle+
  d_{m}^{\dag}|G\rangle-\frac{1+\gamma}{g\sqrt{A}}\sin{\theta_{m}}|e\rangle,
\end{equation}
where
\begin{eqnarray*}
  c_{m}^{\dag} & = &\frac{1}{\sqrt{A}}\sum^{n}_{j=1}\sin (j\theta_{m})
  a_j^{\dagger}, \\
  d_{m}^{\dag} & = & -\frac{\gamma}{\sqrt{A}}\sum^{N}_{j=n+1}
  \sin (j\theta_{m}) a_j^{\dagger},\\
  \gamma & = & n/(N-n+1 ), \\
  A & = & \sqrt{\frac{N+1}{N-n+1} \left[\frac{n}{2} +
      \frac{N+1}{g^2(N-n+1)}
      \sin^2{\theta_{m}})\right]}.
\end{eqnarray*}
\begin{figure}[!htbp]
  \includegraphics[width=8cm]{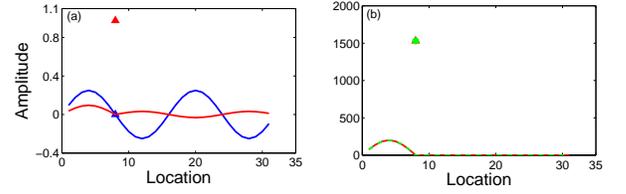}
  \caption{(Color online). (a) Blue (red) solid line and
    triangle each represents the excitation amplitude of the SC and
    the atom as in state $|\psi_{m}\rangle$ ($|\varphi_{m}\rangle $).
    (b) Green dashed (red solid)line and triangle (circle) each represents the excitation
    amplitude of the SC and the atom as in state $|\Psi_{m}\rangle_{SC}$ in the two level
    approximation model (the exact model).
    Here, $g=0.05$ and $m=4$ while all the other parameters are the
    same as before. }
  \label{zero}
\end{figure}

The states $|\psi_{m}\rangle$ and $|\varphi_{m}\rangle$ are demonstrated in
Fig.~\ref{zero}(a). In addition, a simple calculation shows the perfect
reflection appears at $E_{k}=\nu_{m}$. Then the analytical result of
the scattering state within the SC
\begin{equation}
  |\Psi_{m}\rangle_{SC} =C
  \left[\sqrt{\frac{2}{N+1}}\sin(N\theta_{m})c_{m}^{\dag}|G\rangle +
    \frac{\alpha_{2}\sin{\theta_{m}}}{g\sqrt{A}}|e\rangle \right],
\end{equation}
where
\begin{equation*}
  C=-\frac{2i(\gamma+1)\sin{\theta_{m}}}{\eta(\alpha_{1}
    \beta_{2}-\alpha_{2} \beta_{1})}.
\end{equation*}
The state is clearly localized between the atom and the left end of
SC, while the two modes represented by $|\psi_{m}\rangle$ and
$|\varphi_{m}\rangle$ are extended through out the SC (as clearly shown in Fig~\ref{zero}).

\begin{figure}[!htbp]
  \includegraphics[width=8cm]{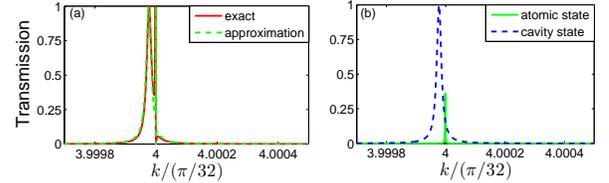}
  \caption{(Color online). Transmission rates vs the incident wave
    vector for different model. (a) The results under the full
    Hamiltonian (red solid line) and two level approximation (green
    dashed line). (b) The results when we only consider one channel which
    is supported by the $m$th eigen-state of the SC (blue dashed line)
    and the atomic state (green solid line).}
  \label{one}
\end{figure}

To test the validness of the two level approximation, we compare the
numerical results in the approximation with those from the exact
model. As shown in Fig.~\ref{one}(a), the transmission spectrum from
the two level approximation agrees well with those from the exact
model, which verifies that the transmission valley is a two level
effect. To further clarify the two level approximation, we give the
transmission spectrum when either one of the two levels is considered,
which is shown in Fig.~\ref{one}(b). Obviously, the transmission
valley just appears in the overlap region due to quantum interference.

\subsection{Numerical results for $\omega_{a}\neq\nu_{m}$}
\label{sec:numer-results-omeg}

\begin{figure}[!htbp]
  \includegraphics[width=8cm]{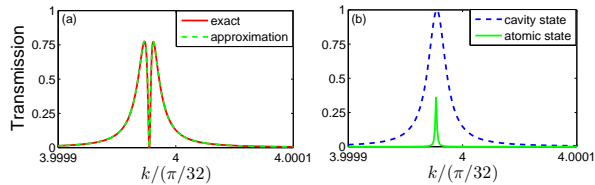}
  \caption{(Color online). Transmission spectrum for various channels.
    The transition frequency of the atom $\omega_{a}=1.847760755\xi$. (a)
    Blue dashed line: the cavity state $|\psi_{m}\rangle$. Green
    solid line: the atomic state $|\varphi_{m}\rangle$. (b) Green dashed
    line: the two level approximation. Red solid line: the exact model.}
  \label{three}
\end{figure}

When the atom is near resonant with the $m$-th mode of the SC, the
analytical expression for the state $|\varphi_{m}\rangle$ is difficult to
obtain. Then we numerically evaluate the state $|\varphi_{m}\rangle$ and
the eigen-energy $\omega_{A}$. The transmission spectrum
within the two level approximation is shown in Fig.~\ref{three}(a).
As before, we also compare these approximate results with those from
the exact model, and find that the approximation is quite accurate.
The transmission spectrum from either one of the two levels are shown
in Fig~\ref{three}(b), which verifies that the transmission valley
comes from the quantum interference between the two channels assisted
by the two levles.

\section{Remarks and Conclusion \label{sec:5}}

In this paper, we have studied the single-photon scattering in a
typical CCA system. Experimentally, the CCA can be realized by the
superconducting transmission line resonators which supports the
single mode microwave electromagnetic field with the resonant
frequency $\omega_c/2\pi\approx3$ GHz~\cite{Wallraff}. The coupling
between neighboring resonators can be realized by via the tunable
capacitances and its strength can be achieved $\xi(\eta)/2\pi=5-100$
MHz~\cite{Wallraff,Mariantoni}. Correspondingly, the two-level atom
can be realized by the superconducting qubit such as flux qubit whose
transition frequency can be tuned by readily adjusting the flux
through the loop, and the coupling strength between the qubit and the
resonator can be achieved $g\approx 0.12 \omega_c$ in a recent
experimental scheme~\cite{TN}.

In conclusion, we propose a simple microscopic model of cavity QED
based on CCA, and prove that the transmission peaks shall appear near
the eigenvalues of the whole SC system. One of the advantages of this
model is that it provides a platform to deal with the non-resonance
modes in the microscopic level. Firstly we show that the SC composed
by a segment of 1D cavity array is a multi-mode cavity by studying the
transmission spectrum through an empty SC. Then we study the
multi-mode effects in the SC system by investigating the transmission
spectrum for the SC interacting with a two-level atom, located at the
antinode or node of a pre-selected resonant mode of SC. When the atom
locates at the antinode, we observe the vacuum Rabi splitting in the
transmission spectrum as expected. However, when the two-level atom
locates at the node of the near resonant mode, we find a valley in the
transmission spectrum, which can be explained by the interference of
the transmission amplitude through two channels, one channel is the
resonant mode, the other is the atomic excited state dressed by the
non-resonance modes. We hope this model can enlighten the study of
multi-mode effects in cavity QED.

\begin{acknowledgments}
  This work is supported by NSF of China (Grant No. 11175247) and
  NKBRSF of China (Grant Nos. 2012CB922104 and 2014CB921202).
\end{acknowledgments}

\section*{APPENDIX: LOCATIONS OF THE PEAKS}
By comparing the coefficients of $\{|j\rangle\}$ and $|e\rangle$ in
the stationary scattering equation \eqref{eq:2}, we obtain
\begin{equation}
  \label{eq:7}
  E_{k} = \omega_{c}+2\xi\cos k,
\end{equation}
and
\begin{eqnarray}
  \label{eq:6}
  \omega_{c}r^{'}+\xi(e^{-ik}+re^{ik})+\eta d_{1} & = & E_{k}r^{'}, \label{a3}\\
  \omega_{c}d_{1}+\eta r^{'}+\xi d_{2} & = & E_{k}d_{1}, \\
  \omega_{c}d_{j}+\xi (d_{j-1}+d_{j+1}) & = & E_{k}d_{j}, \\
  \omega_{c} d_{n}+\xi(d_{n-1}+d_{n+1})+ \lambda g & = & E_{k} d_{n}, \label{a1}\\
  \lambda\omega_{a}+g d_{n} & = & \lambda E_{k}, \label{a2} \\
  \omega_{c}d_{N}+\xi d_{N-1}+t^{'}\eta & = & E_{k}d_{N}, \label{a4}\\
  \omega_{c}t^{'}+\eta d_{N}+\xi t^{'}e^{ik} & = & E_{k}t^{'}. \label{a5}
\end{eqnarray}
where $2 \leq j \leq N-1$ and $j\neq n$.

For $\omega_{a}\neq E_{k}$, Eq.~(\ref{a2}) leads to
\begin{equation}
  \lambda=\frac{g}{E_{k}-\omega_{a}}d_{n}. \label{b1}
\end{equation}
Substituting Eq.~(\ref{b1}) into Eq.~(\ref{a1}), we have
\begin{equation}
  (\omega_{c}+\frac{g^2}{E_{k}-\omega_{a}})d_{n}+\xi (d_{n-1}+d_{n+1})=E_{k}d_{n}.
\end{equation}

We introduce the following parameters:
\begin{eqnarray}
  d_{0} & = & 1+r, \\
  d_{N+1} & = & te^{ik(N+1)}, \\
  \alpha & = & \frac{\omega_{c}-E_{k}}{\xi}=-e^{ik}-e^{-ik},\\
  \beta & = & \frac{\omega_{c}-E_{k}+\frac{g^2}{E_{k}-\omega_{a}}}{\xi}
  =\alpha+\frac{g^2}{\xi(E_{k}-\omega_{a})}, \\
  \gamma & = & \frac{\eta}{\xi}, \\
  \delta & = & e^{ik}-e^{-ik}.
\end{eqnarray}
Then the scattering equation can be expressed in the following matrix
form
\begin{widetext}
  \begin{equation}
    \left(
      \begin{array}{ccccccccccc}
        \alpha+e^{ik} & \gamma & & & & & & & & &\\
        \gamma & \alpha & 1 & & & & & & & &\\
        & 1 & \alpha & 1 & & & & & & &\\
        & & & \ddots & & & & & & &\\
        & & & 1 & \alpha & 1 & & & & &\\
        & & & & 1 & \beta & 1 & & & &\\
        & & & & & 1 & \alpha & 1 & & &\\
        & & & & & & & \ddots & & &\\
        & & & & & & & 1 & \alpha & 1 &\\
        & & & & & & & & 1 & \alpha & \gamma\\
        & & & & & & & & & \gamma & \alpha+e^{ik}
      \end{array}\right)\left(
      \begin{array}{c}
        d_{0}\\
        d_{1}\\
        d_{2}\\
        \vdots\\
        d_{n-1}\\
        d_{n}\\
        d_{n+1}\\
        \vdots\\
        d_{N-1}\\
        d_{N}\\
        d_{N+1}
      \end{array}
    \right) = \left(
      \begin{array}{c}
        \delta\\
        0\\
        0\\
        \vdots\\
        0\\
        0\\
        0\\
        \vdots\\
        0\\
        0\\
        0
      \end{array} \right),
  \end{equation}
\end{widetext}
which is abbreviated as
\begin{equation}
  BD=\Gamma
\end{equation}
with
\begin{equation}
  B=\left(
    \begin{array}{ccc}
      \alpha+e^{ik} & \gamma &\\
      \gamma & A^{n}_{N} & \gamma\\
      & \gamma & \alpha+e^{ik}\\
    \end{array}
  \right) \label{c1}
\end{equation}
and
\begin{equation}
  A^{n}_{N} = \frac {H_{S}-E_{k}} {\xi}.
\end{equation}

By applying the Cramer's Rule, the transmission amplitude is
\begin{equation}
  t=e^{-ik(N+1)}d_{N+1}=e^{-i(k+\pi)(N+1)}\frac{\delta \gamma^2}{|B|}
\end{equation}
with $|B|$ the determinant of the matrix. From Eq.~(\ref{c1}), $|B|$
can be analytical expressed as
\begin{equation}
  |B|=e^{-2ik}|A^{n}_{N}|+e^{-ik}\gamma^{2}(|A^{n-1}_{N-1}|+| A^{n}_{N-1}|)+\gamma^{4}|A^{n-1}_{N-2}|. \label{c2}
\end{equation}

As mentioned before $\eta \ll \xi$, leads to $\gamma \ll 1$. So in
order to gain large transmission probability, $|A^{n}_{N}|$ in the
denominator of Eq.~(\ref{c2}) must be small at least in the order of
$\gamma^{2}$, which clearly shows that the transmission peaks should
be achieved just near the eigenvalues of the SC system. This result
is easy to generalize to multi-atom situation.

\end{document}